\documentclass[copyright]{eptcs}
\providecommand{\event}{J.~Dassow, G.~Pighizzini, B.~Truthe (Eds.): 11th International Workshop\\
on Descriptional Complexity of Formal Systems (DCFS 2009)} 
\providecommand{\volume}{3}
\providecommand{\anno}{2009}
\providecommand{\firstpage}{75} 
\providecommand{\eid}{7}
\usepackage{breakurl}        

\input{dcfs.tex}
\usepackage{multicol}

\newcommand{\bdisplay}{\begin{description}\footnotesize\item[]}
\newcommand{\edisplay}{\end{description}}

\newcommand{\bquot}[1]{\begin{quotation}\small\noindent
  \textbf{#1}\hspace{\labelsep}\ignorespaces}
\newcommand{\equot}{\unskip\end{quotation}}

\usepackage{graphicx} 
\usepackage{algorithm2e}
\SetAlFnt{\small}
\SetCommentSty{text}
\dontprintsemicolon
\usepackage{multirow}

\begin{document}%
     %
\title{Mutation of Directed Graphs -- Corresponding Regular Expressions and 
       Complexity of Their Generation}
\def\titlerunning{Mutation of Directed Graphs -- Corresponding RE and Complexity of Their Generation}

\author{%
\makebox[0pt][c]{Fevzi Belli}\hspace{50mm}\makebox[0pt][c]{Mutlu Beyaz{\i}t}
\institute{
  University of Paderborn\\
  {\small Faculty of Computer Science, Electrical Engineering and Mathematics}\\
  Germany}
\email{\makebox[0pt][c]{belli@adt.upb.de}\hspace{50mm}\makebox[0pt][c]{beyazit@adt.upb.de}}
}

\def\authorrunning{F.~Belli, M.~Beyaz{\i}t}
\maketitle
%
\begin{abstract}
Directed graphs (DG), interpreted as state transition diagrams, are traditionally used 
to represent finite-state automata (FSA). In the context of formal languages, both FSA 
and regular expressions (RE) are equivalent in that they accept and generate, respectively,
type-3 (regular) languages. Based on our previous work, this paper analyzes effects of graph 
manipulations on corresponding RE. In this present, starting stage we assume that the DG 
under consideration contains no cycles. Graph manipulation is performed by deleting or 
inserting of nodes or arcs. Combined and/or multiple application of these basic operators 
enable a great variety of transformations of DG (and corresponding RE) that can be seen as
mutants of the original DG (and corresponding RE). DG are popular for modeling complex 
systems; however they easily become intractable if the system under consideration is 
complex and/or large. In such situations, we propose to switch to corresponding RE in
order to benefit from their compact format for modeling and algebraic operations for 
analysis. The results of the study are of great potential interest to mutation testing.
\end{abstract}

%
%
\section{Introduction and related work}\label{s:introduction}
Most of model-based testing techniques operate on graphs, especially on directed graphs (DG). This has been masterly expressed by one of the testing pioneers, Beizer, as ``Find a graph and cover it!" \cite{Beizer1990, Binder2000}. The basic idea behind ``graph coverage" entails generation of test cases and the selection of a minimum number of them, called ``test suite", in order to cost-effectively exercise a given set of structural or functional issues of the software under test (SUT). A good test coverage increases user confidence in software artifacts, showing that the software is doing everything as it is supposed to do (\textit{positive} testing,~\cite{BL2008}).

For implementation-oriented, white-box testing, nodes of the DG to be covered usually represent the statements of SUT; arcs represent the sequences of those statements \cite{GBBD2005}. For specification-oriented, black-box testing, nodes of the DG may represent the behavioral events of SUT; arcs represent the sequences of those events \cite{Belli2001}.

When using a graph to model of SUT, Belli et al. propose not only to cover the DG model given, but also its complement, showing that the software is \textit{not} doing anything it is \textit{not} supposed to do (\textit{negative} testing, \cite{Belli2001, BL2008}). For this, the authors propose specific manipulation operators of the graph that models SUT. Negative testing approach can be seen in relationship with mutation testing, which is originally a white-box test technique \cite{DLS1978}. Recently, Belli et al. proposed to extend mutation-testing approach to black box, model-based testing \cite{BBW2006}.

A tough problem with complex SUTs is that modeling graphs rapidly become large and thus tedious to work with. If the modeling DG can be interpreted as the transition diagram of a finite-state automaton (FSA), it might be helpful to transform the modeling DG into an algebraic format, i.\,e., regular expressions (RE), and work with this compact formulae instead of spacious graphs (also see \cite{Shaw1980}). Thereby, well-known algorithms can be used to solve the problems concerning the transformation from DG to RE, and v.\,v. \cite{Gill1962, Salomaa1969, HMU2001}. In order to extract the RE from a given DG, one may follow the steps given below:
\begin{itemize}
  \item Convert DG to deterministic FSA (by interpreting the DG as a Moore Machine \cite{Moore1956} and FSA as a Mealy Machine \cite{Mealy1955}).
  \item Convert the FSA to RE by using the widely known algorithms in the literature (also see \cite{HW2007}).
\end{itemize}
In addition, for the opposite chain of transformations, the following steps can be used:
\begin{itemize}
  \item Convert RE to non-deterministic FSA (also see \cite{HSW2001, Geffert2003}).
  \item Convert non-deterministic FSA to a deterministic FSA (and minimize).
  \item Convert the FSA to DG (similar to Mealy - Moore conversion).
\end{itemize}

Application of the basic operators, as introduced in \cite{BBW2006}, to a DG transforms it to another DG, which likely corresponds to a different RE than the original one. Contrarily, the corresponding DG of a manipulated RE differs from the DG that corresponds to the original RE. One of the main objectives of our research is to take the initial steps in order to increase the efficiency of mutation testing by determining, if possible, correspondences between DG and RE modifications. In testing literature, there are many varied constructs, such as DG, FSA, EFSA (extended FSA), ESG and state charts etc., which are used to model a SUT. Each of these graph-based representations possesses different syntax and semantics. In fact, in many cases they are presented as an extension of one another. The common arguments which can be drawn on these structures are (1) they all have (extended) RE counterparts, and (2) the more complex the SUT gets the harder they are to work with in their graphic format. \cite{Myhill1957, Salomaa1969, Belli2001, BBW2006}

To our knowledge there is no approach which aims to manipulate the corresponding RE in order to reflect alterations of the mutation operators performed on the given DG, or v.\,v. However, it is worth mentioning that there are several works on the algebra of RE which enables the transformations via some defined system of rules, such as \cite{SH1963, Brzozowski1964, Salomaa1966}. Taken this into account and based on DG and RE, the next section introduces the notions used in this paper, defines basic operators for graph manipulations and finally introduces the ``sum of products" format for canonic representation of regular expressions. Section \ref{s:approach} applies those basic operators to DG and algorithmically generates their corresponding RE. Complexity of these algorithms are determined (see also the Appendix), before Section \ref{s:conclusion} concludes the paper with a summary of results already achieved and research work planned.
\section{Notions used}\label{s:notions}
This section briefly and semi-formally summarizes notions we need to launch the discussion 
in Section~\ref{s:approach}.
\subsection{Directed graphs and regular expressions}
\begin{definition}\label{def:dg}
A \textit{directed graph} (\textit{DG}) is the tuple $(V, A)$ where $V$ is a finite set 
of \textit{nodes}, i.\,e.,\linebreak \hbox{$V= \{v_1, \ldots, v_n\}$}, and $A$ is a finite set 
of \textit{directed arcs} which are ordered pairs of elements of $V$, i.\,e., 
\hbox{$A = \{a_1, \ldots, a_m\} \subseteq V \times V$}, where each $a_i = (v_j, v_k)$ for 
some $j$, $k$.
\end{definition}

\begin{definition}\label{def:re}
A \textit{regular expression} (\textit{RE}) consists of \textit{symbols} of an \textit{alphabet} and is used to express a set of \textit{strings} (or \textit{words}), i.\,e., a \textit{language}. In an operational perspective, a RE can be assumed to be a sequence of symbols $a, b, c, \ldots$ of an alphabet which can be connected by operations
\begin{itemize}
  \item \textit{sequence} (``$.$", but usually no explicit operation symbol, e.\,g., ``$ab$" means ``$b$ follows $a$"),
  \item \textit{selection} (``$+$", e.\,g. ``$a+b$" means ``$a$ or $b$"),
  \item \textit{iteration} (``$^*$", \textit{Kleene's Star Operation}, e.\,g.,
  \begin{itemize}
    \item[--] ``$a^*$" means ``$a$ will be repeated arbitrarily";
    \item[--] ``$a^+$" means at least one occurrence of ``$a$").
  \end{itemize}
\end{itemize}
These operations are also applied on RE other than simple symbols and, as usual, \textit{parenthesization} is used to guarantee the intended precedence and associativity.
\end{definition}

\begin{figure}
  \centering
  \includegraphics[scale=0.70]{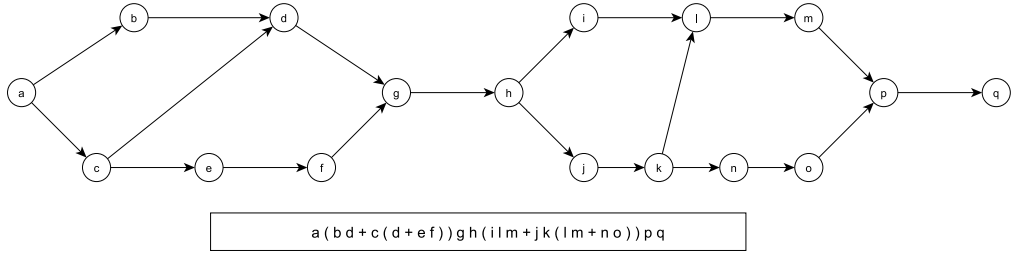}
  \caption{A sample DG and its corresponding RE}
  \label{fig:sample_dg}
\end{figure}

A sample DG and its corresponding RE are given in Figure \ref{fig:sample_dg}. In order to define a RE representation of a DG, we need to distinguish some nodes as \textit{start} nodes and some others as \textit{finish} nodes. In this context, the set of nodes is considered as the \textit{alphabet} (the set of \textit{symbols}) and the \textit{words} (\textit{strings}) in the language expressed by the RE are, in fact, the node sequences forming paths connecting start nodes to finish nodes in the graph. This convention has been introduced in \cite{Belli2001} to define \textit{event sequence graphs} (\textit{ESG}).
\subsection{Operators for manipulation of directed graphs}
For manipulation of a graph, or a DG, elementary operations can be classified under two categories, \textit{insertion} $(i)$ and \textit{omission} $(o)$, and since a DG consists of nodes and edges, the manipulation operators can be specified as \textit{node insertion}~$(i_n)$, \textit{node omission} $(o_n)$, \textit{arc insertion} $(i_a)$ and \textit{arc omission} $(o_a)$ operators.

\begin{definition}\label{def:mo}
DG manipulation operators transform a DG to another DG and defined as follows:
\begin{itemize}
  \item \textit{Arc insertion} operator adds a new arc $(v_j, v_k)$, where $v_j, v_k \in V$, to the DG $(V, A)$:
    \begin{center}
      $(v_j, v_k) i_a : (V, A) \to (V, A \cup \{(v_j, v_k)\})$.
    \end{center}
  \item \textit{Arc omission} operator deletes an arc $(v_j, v_k)$, where $v_j, v_k \in V$, from the DG $(V, A)$:
    \begin{center}
      $(v_j, v_k) o_a : (V, A) \to (V, A \setminus \{(v_j, v_k)\})$.
    \end{center}
    It is possible that some nodes are left with no ingoing and/or outgoing arcs.
  \item \textit{Node insertion} operator adds a new node $v \notin V$ to the DG $(V, A)$
  together with possibly nonzero number of arcs $\{a_1, \ldots, a_k\}$ connecting this node to the remaining nodes:
    \begin{center}
      $(v, {a_1, \ldots, a_k}) i_n : (V, A) \to (V \cup {v}, A \cup \{a_1, \ldots, a_k\})$.
    \end{center}
  \item \textit{Node omission} operator deletes a node $v \in V$ from the DG $(V, A)$
  together with all the arcs ${a_1, \ldots, a_k}$ A ingoing to and outgoing from the deleted node:
    \begin{center}
      $(v) o_n : (V, A) \to  (V \setminus \{v\}, A \setminus \{a_1, \ldots, a_k\})$.
    \end{center}
\end{itemize}
\end{definition}

\begin{figure}
  \centering
  \includegraphics[scale=0.70]{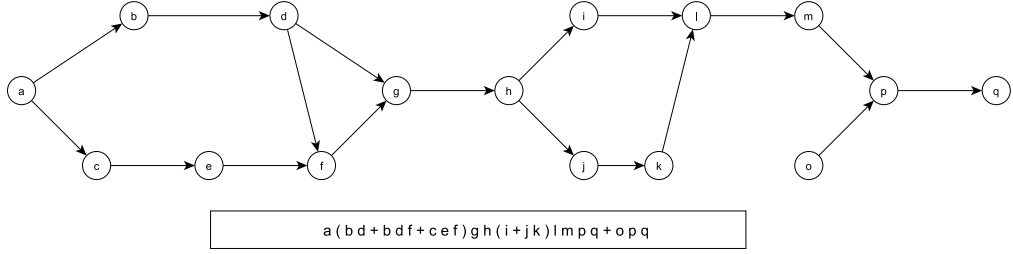}
  \caption{The manipulated DG and RE by application of $(cd)o_a (df)i_a (n)o_n$}
  \label{fig:sample_dg_2}
\end{figure}

Figure \ref{fig:sample_dg_2} results from the application of basic manipulation operators to the DG in Figure \ref{fig:sample_dg}.

\subsection{Sum of products format for RE and auxiliary functions}
In order to carry out transformation on the RE in an algorithmic way, we introduce, in analogy to Boolean Algebra, a canonical representation for RE under consideration and some auxiliary functions which operate on RE.

\begin{definition}
A given RE is in the \textit{sum of products format} (\textit{SOPF}), if it is represented as the sum of finitely many \textit{product terms}, each of which is in one of the following forms:
\begin{itemize}
  \item $r$
  \item $R^*$
  \item Finite concatenations of $r$ and/or $R^*$ (such as $rR^*$, $R^*r$ and $rR^*rR^*$, etc.)
\end{itemize}
where $r$ is an arbitrary finite string (formed by only the concatenation of symbols) and $R$ is a RE in SOPF. Note that the SOPF is a very simple and straightforward format which highly disregards the compactness.
\end{definition}
\begin{example} Let the RE in Figure \ref{fig:sample_dg} be R, then SOPF of R is given as below
\begin{align*}
SOPF(R)=\,& abdghilmpq + abdghjklmpq + abdghjknopq +{} \\
          & acdghilmpq + acdghjklmpq + acdghjknopq +{} \\
          & acefghilmpq + acefghjklmpq + acefghjknopq.\tag*{$\diamond$}
\end{align*}
\end{example}

\begin{definition}
Let $D$ be the DG with the set of vertices $\{v_1, \ldots, v_n\}$ and $R$ be the corresponding RE, then we define:
\begin{itemize}
  \item $pt(R, s)$ to be the set of product terms which contain the string $s$,
  \item $ht(P, s)$ to be the set of product terms which are the beginning (\textit{head}) subterms, ending with the first occurrence of the string $s$, of the product terms in the set $P$, and
  \item $tt(P, s)$ to be the set of product terms which are the ending (\textit{tail}) subterms, beginning with the last occurrence of the string $s$, of the product terms in the set $P$.
\end{itemize}
\end{definition}

\begin{example} Let SOPF of the RE in Figure \ref{fig:sample_dg} be R, then we have
\begin{itemize}
  \item $pt(R, jkl) = \{abdghjklmpq, acdghjklmpq, acefghjklmpq\}$,
  \item $ht(P, f) = \{acef\}$, and
  \item $tt(P, gh) = \{ghilmpq, ghjklmpq, ghjknopq\}$.\hfill$\diamond$
\end{itemize}
\end{example}
\section{Approach: graph manipulation and effects on corresponding regular expression}\label{s:approach}
As basically discussed in the introduction section, the problem, in its general form, is to expose the underlying correspondences between DG and RE manipulations. More precisely, we have following situation: Given a DG and the corresponding RE, we want to reflect the result of DG transformations stemming from applications of basic manipulation operators to the corresponding RE, and v.\,v. For this purpose, we assume that (1) the initial and transformed DG have no cycles and (2) all the RE are in the SOPF. Furthermore, the analysis of the effects of RE manipulation on corresponding DG, is postponed; in this stage we just focus on the effects of manipulating DG on its RE.

Under the assumptions stated above, the following subsections outline straightforward 
algorithms for basic manipulation of DG by transforming its corresponding RE, and, in the 
discussion ahead, $|P|$ and $|p|$ are defined to be upper bounds on the number of product 
terms and on the lengths of the product terms in the given RE. Complexity values of the 
auxiliary algorithms are included in Appendix and Table~\ref{tab:auxcomp}; they are 
necessary for the validation of the worst case time complexity results of the DG 
manipulation algorithms which are analyzed in the next subsections.
\subsection{Arc operators}
Following, omission and insertion operations are applied to arcs.
\subsubsection{Arc insertion}
Algorithm \ref{alg:ai} outlines the addition of new paths connecting start and finish nodes in the DG as product terms to the given RE, during the insertion of the arc $(v_i, v_j)$, where $v_i, v_j \in V$, to the DG. During the insertion, no product term in the RE should contain the symbol $v_j$ before $v_i$. Otherwise, the operation produces a cycle.

\begin{algorithm}
\KwIn{$R$ - a RE in SOPF, \;\Indp\Indp\Indp
        $(v_i, v_j)$, where $v_i, v_j \in V$ - arc to be inserted}
\KwOut{$R$ - RE is updated with the insertion of the arc $(v_i, v_j)$ in SOPF}
\BlankLine
\Indp
  $A = pt(R, v_i)$  \tcp{Find the set of product terms containing $v_i$}\;
  $B = pt(R, v_j)$  \tcp{Find the set of product terms containing $v_j$}\;
  $A' = ht(A, v_i)$ \tcp{Construct the set of head subterms for $A$}\;
  $B' = tt(B, v_j)$ \tcp{Construct the set of tail subterms for $B$}\;
  $C' = A'.B'$      \tcp{Perform the set concatenation operation on $A'$ and $B'$}\;
  $R = R + RE(C')$  \tcp{Add the terms in $C'$ as product terms to $R$}\;
\Indm
\caption{Arc Insertion}
\label{alg:ai}
\end{algorithm}

As implied by Algorithm \ref{alg:ai}, in the insertion of the arc $(v_i, v_j)$, the number of new product terms to be added to the RE is given by $|A'||B'|$, where $|A'|$ is the number of (distinct) head subterms leading to the node $v_i$ from the start nodes and $|B'|$ is the number of (distinct) tail subterms leading to the finish nodes from the node $v_j$.

Algorithm \ref{alg:ai} is terminating, since all the subroutines are executed in finite time. Furthermore, a straightforward calculation using the values in Table \ref{tab:auxcomp} shows that Algorithm \ref{alg:ai} has the worst case time complexity $O(|P|^4 |p|)$. It is possible to reduce this complexity value to $O(|P|^2 |p|)$ by performing the set concatenation without filtering the duplicate product terms while constructing the set $C'$ in $O(|P|^2 |p|)$ time. These duplicate terms can be left out during the set union operation without affecting its worst case time complexity.
\subsubsection{Arc omission}
Omission of an arc may leave some nodes with no ingoing and/or outgoing edges. These nodes are considered as valid start and/or finish nodes respectively, because the succeeding operations may introduce new edges to such nodes. Thus, Algorithm \ref{alg:ao} updates the given corresponding RE after the omission of the arc $(v_i, v_j)$, where $v_i, v_j \in V$, from the DG.

\begin{algorithm}
\KwIn{$R$ - a RE in SOPF, \;\Indp\Indp\Indp
        $(v_i, v_j)$, where $v_i, v_j \in V$ - arc to be omitted}
\KwOut{$R$ - RE is updated with the omission of the arc $(v_i, v_j)$ in SOPF}
\BlankLine
\Indp
  $A = pt(R, v_i)$      \tcp{Find the set of product terms containing $v_i$}\;
  $B = pt(R, v_j)$      \tcp{Find the set of product terms containing $v_j$}\;
  $C = pt(R, v_iv_j)$   \tcp{Find the set of product terms containing $v_iv_j$}\;
  $A' = \emptyset$\;
  \textbf{if} $A = C$ \textbf{then} \\
  $\quad A' = ht(A, v_i)$   \tcp{Construct the set of head subterms for $A$}\;
  \textbf{endif} \\
  $B' = \emptyset$\;
  \textbf{if} $B = C$ \textbf{then} \\
  $\quad B' = tt(B, v_j)$ 	\tcp{Construct the set of tail subterms for $B$}\;
  \textbf{endif} \\
  $C' = A' \cup B'$     \tcp{Union of $A'$ and $B'$}\;
  $R = R - RE(C)$       \tcp{Remove the product terms in $C$ from $R$}\;
  $R = R + RE(C')$      \tcp{Add the terms in $C'$ as product terms to $R$}\;
\Indm
\caption{Arc Omission}
\label{alg:ao}
\end{algorithm}

In Algorithm \ref{alg:ao}, the number of product terms to be added to and removed from 
the RE is given by $|A'|+|B'|$ and $|C|$, respectively, where $|A'|$ is the number 
of (distinct) head subterms leading to the node~$v_i$ from the start nodes, $|B'|$ is the 
number of (distinct) tail subterms leading to the finish nodes from the node $v_j$ and $|C|$ 
is the number of (distinct) product terms containing the sequence $v_iv_j$.

The worst case time complexity of Algorithm \ref{alg:ao} is $O(|P|^2 |p|)$ (see Table \ref{tab:auxcomp} for the complexity of auxiliary algorithms), and it runs in finite time.
\subsection{Node operators}
As a next step, omission and insertion operations are applied to nodes.
\subsubsection{Node insertion}
Node insertion is a higher level operation when compared to arc manipulation operations, because it generally requires connecting the node to the remaining nodes. To do this, first, the inserted node is considered as a valid start and finish node. Later, the following arc insertions take place. Accordingly, Algorithm \ref{alg:ni} can be applied to update the corresponding RE with the insertion of the node~$v_i$ together with the arcs $(v_i, x_j)$ and $(y_k, v_i)$, where $v_i \notin V$ and $x_j, y_k \in V$ for $j = 1, \ldots, s$ and $k = 1, \ldots, t$, to the DG.

\begin{algorithm}
\KwIn{$R$ - a RE in SOPF, \;\Indp\Indp\Indp
       $v_i \notin V$ - node to be inserted \;
       $(v_i, x_j)$, where $x_j \in V$, $j = 1, \ldots, s$ - outgoing arcs to be inserted\;
       $(y_k, v_i)$, where $y_k \in V$, $k = 1, \ldots, t$ - ingoing arcs to be inserted}
\KwOut{$R$ - RE is updated with the insertion of the node $v_i$ in SOPF}
\BlankLine
\Indp
 $R = R + v_i$     \tcp{Add the symbol $v_i$ as a product term to $R$}\;
 \textbf{for each} $(v_i, x_j)$ \textbf{do} \\\Indp
   insert the arc $(v_i, x_j)$ and update $R$     \tcp{See Algorithm \ref{alg:ai}}\;\Indm
 \textbf{endfor} \\
 \textbf{for each} $(y_k, v_i)$ \textbf{do} \\\Indp
   insert the arc $(y_k, v_i)$ and update $R$     \tcp{See Algorithm \ref{alg:ai}}\;\Indm
 \textbf{endfor} \\
 \textbf{if} $s \geq 1$ \textbf{or} $t \geq 1$ \textbf{then} \\\Indp
       $R = R - v_i$     \tcp{Remove the product term $v_i$ from $R$}\;\Indm
 \textbf{endif} \\
\Indm
\caption{Node Insertion}
\label{alg:ni}
\end{algorithm}

It is straightforward to note that, given the set union and arc insertion operations run in finite time, Algorithm \ref{alg:ni} runs in finite time, and it has \hbox{$O((s+t) (|P|^2 |p|))$} worst case time complexity where $s$ and~$t$ are the number of ingoing and outgoing arcs to be inserted, respectively. Note that, in a DG with no cycles, $(s+t) \leq n$ always holds and $|p|$ can be chosen to be $n$.
\subsubsection{Node omission}
Node omission entails the deletion of the node and the arcs related to it, and therefore is also a higher level operation with respect to the arc manipulation operations. For omission of a node, the node is disconnected from the rest of the graph and considered as a valid start and finish node, and later removed. Algorithm \ref{alg:no} shows the steps to update the corresponding RE with the omission of the node $v_i$ (and all the arcs $(v_i, x_j)$ and $(y_k, v_i)$, where $x_j, y_k \in V$) from the~DG.

\begin{algorithm}
\KwIn{$R$ - a RE in SOPF, \;\Indp\Indp\Indp
        $v_i \in V$ - node to be omitted \;
        $(v_i, x_j)$, where $x_j \in V$, $j = 1, \ldots, s$ - outgoing arcs to be omitted \;
        $(y_k, v_i)$, where $y_k \in V$, $k = 1, \ldots, t$ - ingoing arcs to be omitted}
\KwOut{$R$ - RE is updated with the omission of the node $v_i$ in SOPF}
\BlankLine
\Indp
  \textbf{for each} $(v_i, x_j)$ \textbf{do} \\\Indp
    omit the arc $(v_i, x_j)$ and update $R$     \tcp{See Algorithm \ref{alg:ao}}\;\Indm
  \textbf{endfor} \\
  \textbf{for each} $(y_k, v_i)$ \textbf{do} \\\Indp
    omit the arc $(y_k, v_i)$ and update $R$     \tcp{See Algorithm \ref{alg:ao}}\;\Indm
  \textbf{endfor} \\
  $R = R - v_i$     \tcp{Remove the product term $v_i$ from $R$}\;
\Indm
\caption{Node Omission}
\label{alg:no}
\end{algorithm}

In Algorithm \ref{alg:no}, the operations arc omission and set difference takes finite
number of steps to complete. Furthermore, since the DG has no cycles, the loops are 
executed at most $n$ times. Thus, the algorithm runs in finite time. In addition, in the 
worst case, running time complexity of Algorithm \ref{alg:no} is $O(k |P|^2 |p|)$ 
where~$k$ is the total number of arcs to be omitted. Also, in a DG with no cycles, 
$k < n$ always holds and choosing $|p| = n$ is valid.
\section{Conclusion and future work}\label{s:conclusion}
This paper considers the effects of basic DG manipulations on the corresponding~RE
and outlines algorithms in order to transform the RE accordingly, where DG contains no 
cycles. Hence, it is an initial step to lay out the correspondence between DG and RE 
mutations from a practical point of view. Some of the main implications of the study,
so far, can be summarized as below in two parts:

(i) Format of the RE: The size of a RE can be defined as its length, i.\,e., the total number of symbols and operators in the RE, and is determined by its format. The size, thus the format, of the RE has a direct effect on the efficiency of the operations. Unfortunately, SOPF is a kind of ``worst-case" format where the compactness is not a concern. However, it helps to keep the algorithms straightforward and simple, and it seems easier to conserve since no additional transformations are required to preserve the format of the RE. Nevertheless, the derived complexity values should be interpreted as the ``worst" of the worst case time complexity values (keeping in mind that this does not always lead to worst performance in practice).

(ii) Extent of the approach: The DG in our present paper are assumed to be free of cycles, but this does not necessarily mean that the DG models which contain cycles are completely out of the scope. One can apply different cycle omission strategies, such as traveling cycles at most a predefined number of times, in combination with the underlying semantics of the system and the indexing mechanism to update or ``flatten" the DG model. Inevitably, the resulting model is only a submodel, however, in practice, there might cases where it is preferable.

On the other hand, our future work will include DG with cycles and enhance the format of the RE without sacrificing the (practical) efficiency which might stem from possible additional transformations. It is one of our concerns to improve the compactness of the RE by keeping it in another format (like perhaps product of sums format (POSF), which seems to be somehow more promising, etc.). However, it would be better and nicer to develop an approach which handles the manipulation operators in an algebraic manner without any respect to the format of the RE.
\bibliographystyle{eptcs}
\bibliography{beyazit}

%
\section*{Appendix. Some auxiliary functions and their complexity}
Worst case time complexity values of some related auxiliary functions are given in Table \ref{tab:auxcomp}. In order to interpret the complexity values correctly, note following:
\begin{itemize}
  \item $|P|$ is an upper bound on the number of product terms in $P$, i.\,e., the number of product terms in $P$, and $|p|$ is an upper bound on lengths of the product terms in $P$, i.\,e., the length of the longest product term in $P$.
  \item $|p'|$ is the length of the product term $p'$.
  \item $|s|$ is the length of the string $s$.
\end{itemize}
Note that the sets A, B and C, and the RE R are also sets of product terms.

\renewcommand{\arraystretch}{1.4}
\begin{table}[h]
\caption{Worst Case Time Complexity Values for Some Auxiliary Functions}\label{tab:auxcomp}
  \begin{center}
  \begin{tabular}[c]{|l|c|}
    \hline
    \rule[-1.5mm]{0mm}{6mm}\makebox[10cm]{\textbf{Function}} & \makebox[2cm]{\textbf{Complexity}} \\
    \hline
    Removal of a product term from a set: $P = P - p'$ & $O(|P|(|p|+|p'|))$ \\\hline
    Addition of a product term to a set: $P = P + p'$ & $O(|P|(|p|+|p'|))$ \\\hline
    Set Union: $C = A \cup B$ & $O(|A||B| (|a|+|b|))$ \\\hline
    Set Concatenation: $C = A . B$ & $O((|A||B|)^2 (|a|+|b|))$ \\\hline
    Extraction of Tail Product Terms: $tt(P, s)$ where $|s| = 1,2$ & $O(|P|^2 |p|)$ \\\hline
    Extraction of Head Product Terms: $ht(P, s)$ where $|s| = 1,2$ & $O(|P|^2 |p|)$ \\\hline
    Extraction of Product Terms: $pt(R, s)$ where $|s| = 1,2$ & $O(|P|^2 |p|)$ \\\hline
  \end{tabular}
  \end{center}
\end{table}
%
\end{document}